\documentclass[a4paper,11pt]{article}
\pdfoutput=1 

\usepackage{jinstpub} 
\usepackage{subfigure}

\title{Performance of irradiated thin n-in-p planar pixel sensors for the ATLAS Inner Tracker upgrade}

\author[1,a]{N. Savi\'{c},\note{Corresponding author.}}
\author[a]{J. Beyer,}
\author[b]{B. Hiti,}
\author[b]{G. Kramberger,}
\author[a]{A. La Rosa,}
\author[a]{A. Macchiolo,}
\author[b]{I. Mandi\'{c},}
\author[a]{R. Nisius,}
\author[b]{M. Petek}

\affiliation[a]{Max-Planck-Institut f\"ur Physik (Werner-Heisenberg-Institut),\\ F\"ohringer Ring 6, 80805 M\"unchen, Germany}
\affiliation[b]{Jo\v{z}ef Stefan Institut,\\ Jamova 39, 1000 Ljubljana, Slovenia}

\emailAdd{Natascha.Savic@mpp.mpg.de}

\abstract{The ATLAS collaboration will replace its tracking detector with new all silicon pixel and strip systems. This will allow to cope with the higher radiation and occupancy levels expected after the 5-fold increase in the luminosity of the LHC accelerator complex (HL-LHC). In the new tracking detector (ITk) pixel modules with increased granularity will implement to maintain the occupancy with a higher track density. In addition, both sensors and read-out chips composing the hybrid modules will be produced employing more radiation hard technologies with respect to the present pixel detector. Due to their outstanding performance in terms of radiation hardness, thin n-in-p sensors are promising candidates to instrument a section of the new pixel system. Recently produced and developed sensors of new designs will be presented. To test the sensors before interconnection to chips, a punch-through biasing structure has been implemented. Its design has been optimized to decrease the possible tracking efficiency losses observed. After irradiation, they were caused by the punch-through biasing structure. A sensor compatible with the ATLAS FE-I4 chip with a pixel size of 50x250\,$\mathrm{\mu}$m$^{2}$, subdivided into smaller pixel implants of 30x30\,$\mathrm{\mu}$m$^{2}$ size was designed to investigate the performance of the 50x50\,$\mathrm{\mu}$m$^{2}$ pixel cells foreseen for the HL-LHC. Results on sensor performance of 50x250 and 50x50\,$\mathrm{\mu}$m$^{2}$ pixel cells in terms of efficiency, charge collection and electric field properties are obtained with beam tests and the Transient Current Technique.}


\proceeding{18$^{\text{th}}$ International Workshop on Radiation Imaging Detectors\\
	2-6 July 2017\\
	Krakow, Poland}

\begin{document}
	\maketitle
	\flushbottom
	
	\section{ATLAS pixel detector upgrade for HL-LHC}
	In view of the high luminosity phase of the LHC (HL-LHC) starting around 2025 the ATLAS experiment will undergo a major upgrade of its tracker system. The ATLAS pixel system is expected to be exposed to fluences up to 1.4$\times$10$^{16}$ $\mathrm{n}_{\mathrm{eq}}/\mathrm{cm}^2$ (1 MeV neutron equivalent) \cite{atlasUp} taking into account one replacement and including a safety factor of 1.5. To be able to maintain the performance in the new and more challenging environment different technologies are developed to  improve radiation hardness and maintain a similar occupancy as in Run 2 despite a higher track density. The baseline technology for all pixel layers except the innermost one is based on pixel modules assembled with thin planar n-in-p sensors. The modules investigated in this paper consist of 100\,$\mathrm{\mu}$m and 150\,$\mathrm{\mu}$m thick sensors produced at CiS (Germany). The sensors are interconnected with solder bump-bonding to ATLAS FE-I4 read-out chips. Bare sensors are characterized by means of the Edge Transient Current Technique (Edge-TCT) at the Jo\v{z}ef Stefan Institut allowing to study the profiles of drift velocity, charge collection and electric field inside heaviliy irradiated silicon detectors to understand the charge transport process in these devices. Sensors from the same production are assembled into modules measured with particle beams at the CERN-SpS. The results of Edge-TCT measurements are discussed for thin either unirradiated or irradiated sensors up to an irradiation level of 1$\times$10$^{16}$ $\mathrm{n}_{\mathrm{eq}}/\mathrm{cm}^2$. Previous results on the hit efficiency and power dissipation revealed that modules with thin sensors are a promising candidate for the innermost layers \cite{nani2}. Given the high particle rate expected at HL-LHC smaller pixel dimensions with respect to the ones currently implemented in the FE-I3 chip (50x400\,$\mathrm{\mu}$m$^{2}$) \cite{fei3} and the FE-I4 chip (50x250\,$\mathrm{\mu}$m$^{2}$) \cite{fei4} are necessary to keep the occupancy at an acceptable level. The new read-out chip for the future ATLAS pixel system with a pixel cell of 50x50\,$\mathrm{\mu}$m$^{2}$ produced in 65\,nm CMOS technology is being developed by the CERN RD53 Collaboration \cite{rd53} and the first prototype, the RD53A chip, is foreseen to be ready by the end of 2017 \cite{chip}. At CiS sensors were produced with 50x50\,$\mathrm{\mu}$m$^{2}$ pixel cells, connected by a metal layer in such a way to be still compatible with the FE-I4 chip footprint. Results on the in-pixel efficiency of small pixel implants before irradiation were previously shown in \cite{nani2} and will now be shown after an irradiation up to 5$\times$10$^{15}$ $\mathrm{n}_{\mathrm{eq}}/\mathrm{cm}^2$.
	
	\section{Characterization of thin n-in-p planar pixel sensors}
A production of thin planar n-in-p pixel on 4-inch wafers of p-type FZ silicon has been recently completed at CiS. The wafers are locally thinned below each single structure from a starting thickness of 500\,$\mathrm{\mu}$m to an active thickness of 100\,$\mathrm{\mu}$m or 150\,$\mathrm{\mu}$m by using anisotropic KOH etching creating backside cavities in the wafer. This technology does not require handle wafers during the thinning process, thus it is potentially cheaper than the alternative method employing SOI wafers. A more detailed description of the production process is given in Ref.~\cite{nani}. The standard FE-I4 pixel pitch with a 50x250\,$\mathrm{\mu}$m$^{2}$ cell was implemented together with the common punch-through design which was found to retain the highest efficiencies after irradiation \cite{nani}.

	\subsection{The Edge Transient Current Technique}
	The Transient Current Technique with the laser beam pointing at the sensor edge is a powerful method to get insights into the charge collection and electric field properties across the sensor depth. The extraction of the charge collection and velocity profile is based on the measurement of time evolution of the induced current pulse. The technique is known as Edge-TCT and a detailed description can be found in Ref.~\cite{Gregor}. In Edge-TCT an infra-red laser with an approximate beam FWHM of 8-12\,$\mathrm{\mu}$m and a long penetration depth in silicon of over 1~mm is shining on the edge of the device, a thin bare sensor of 100\,$\mathrm{\mu}$m or 150\,$\mathrm{\mu}$m thickness, and scanned along the sensor depth and the long direction of the pixel. Charge carriers, electron/hole pairs, are released inside the sensor traversing to their corresponding electrodes. An electrical signal created by the movement of the charge carriers is amplified with an external current amplifier. 
	
		\begin{figure}[h!]
		\centering     
		\subfigure[]{\label{fig:a}\includegraphics[width=80mm, height=65mm]{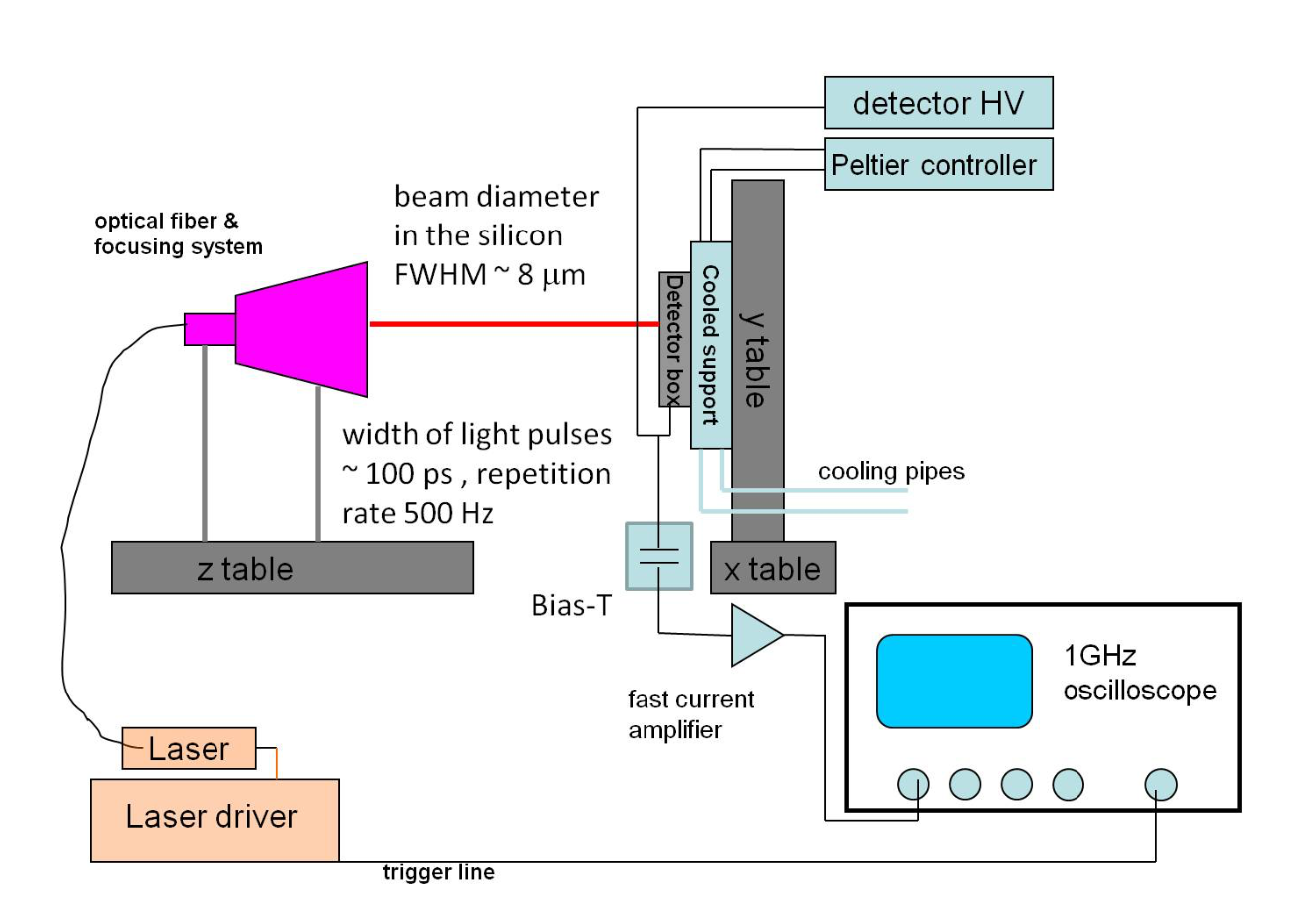}}
		\subfigure[]{\label{fig:b}\includegraphics[width=45mm]{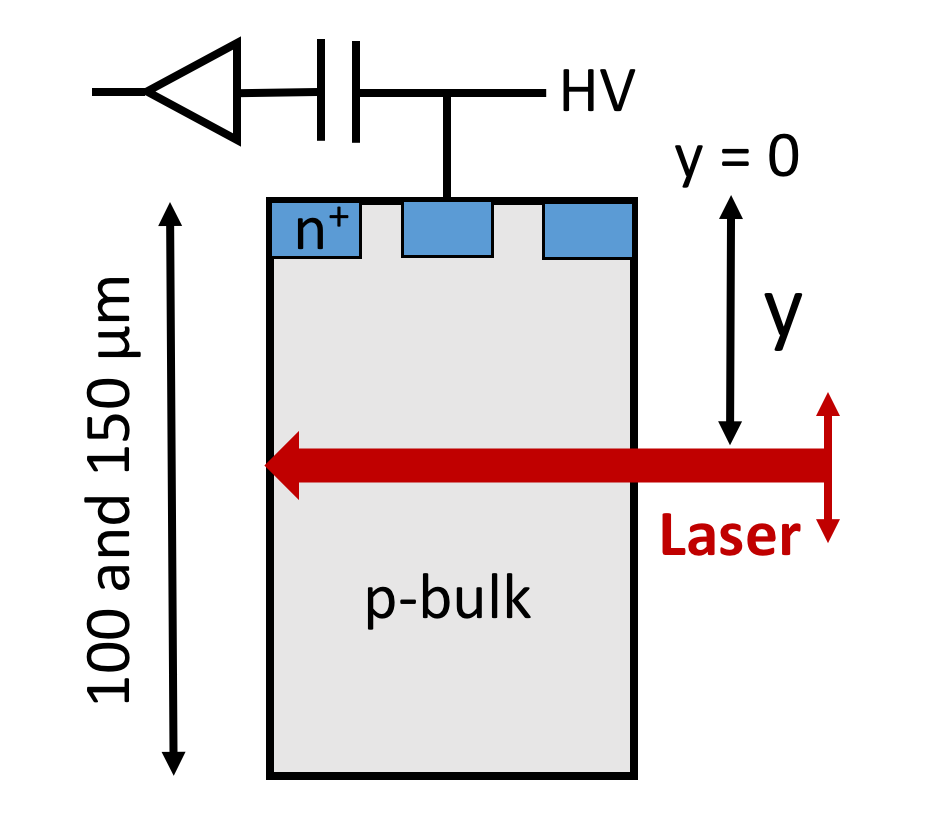}}
		\caption{Outlined description of the Edge-TCT set-up from \textit{Particulars} \cite{particulars}: (a) schematics of the experimental set-up \cite{Igor} and (b) detector side view with the laser beam penetrating at the detector edge.}
		\label{TCTsetup}
	\end{figure}

	\noindent\hspace*{0mm}All pixels on the frontside are connected to high voltage via a wire bond attached to the bias ring surrounding the pixels, while the backside is set to ground. One single pixel is connected to readout via an additional wire bond. The device is mounted onto a copper support, which is thermally stabilized at room temperature in the case of no irradiation and cooled to -20$^{\circ}$C via a Peltier Cooler when irradiated. The schematics of the Edge-TCT measurement set-up is shown in Fig.~\ref{TCTsetup}.

	\subsection{Charge collection properties}
\begin{figure}[h!]
		\centering     
		\subfigure[]{\label{fig:a}\includegraphics[width=65mm, height=47mm]{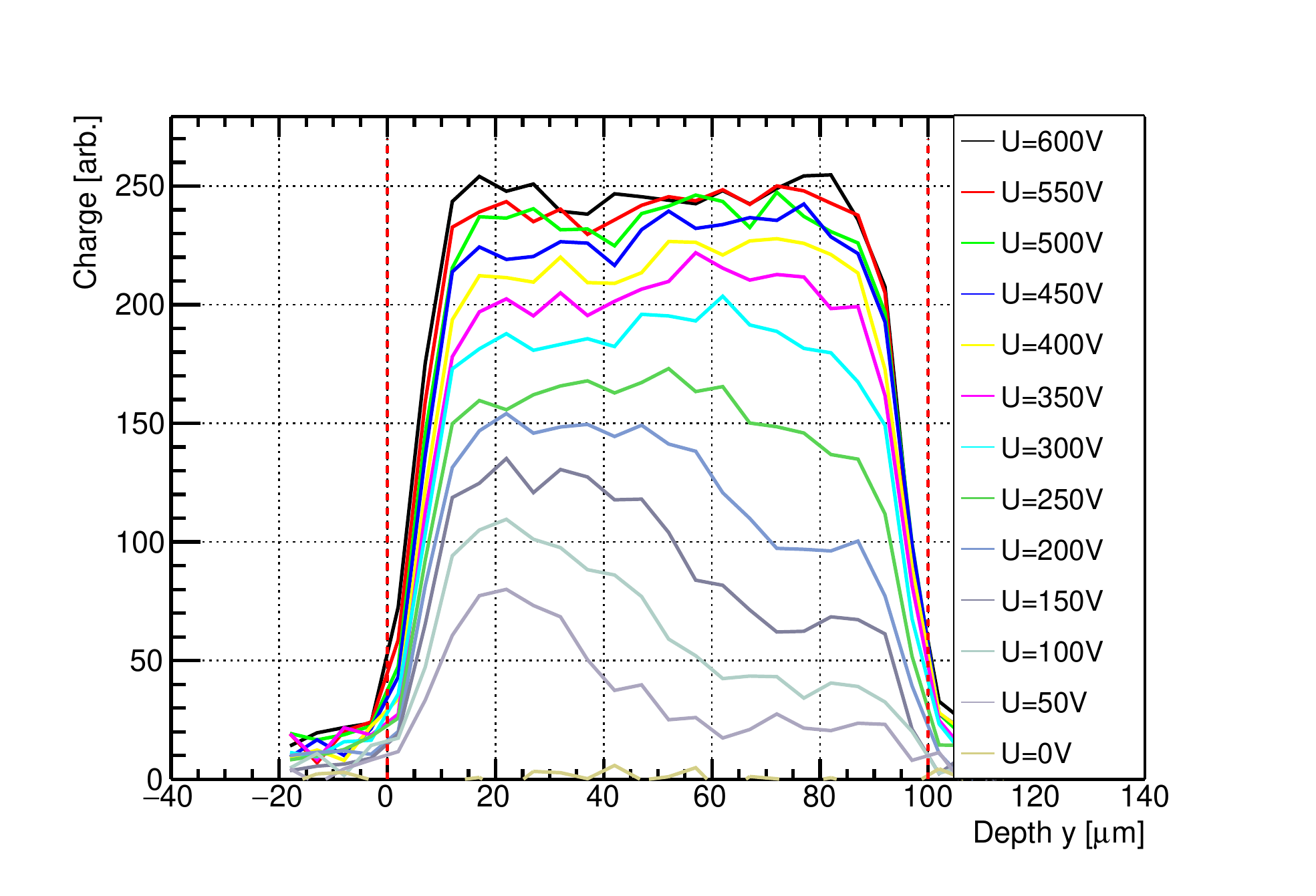}}
		\subfigure[]{\label{fig:b}\includegraphics[width=65mm]{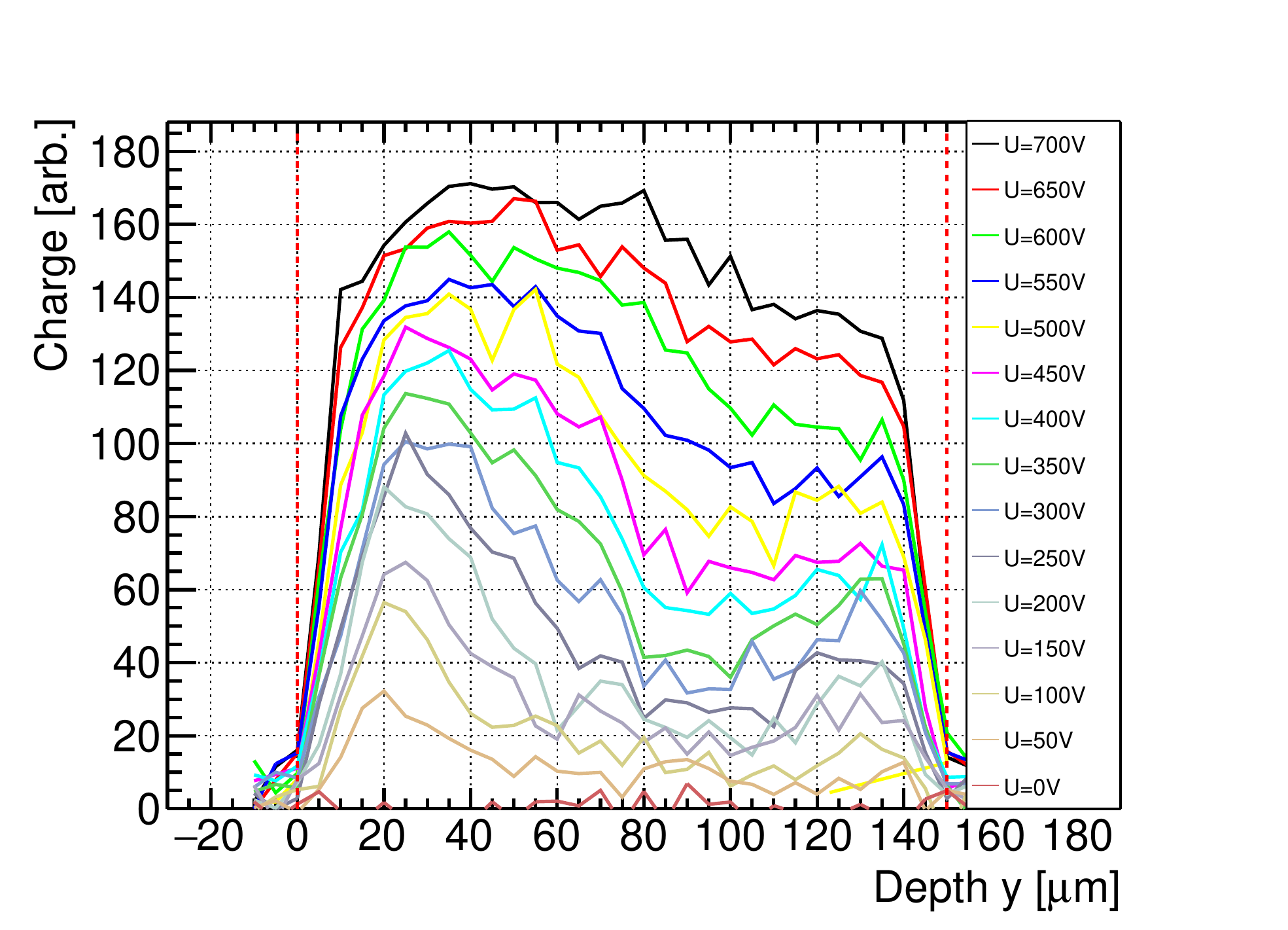}}
		\caption{The charge collection as a function of sensor depth is shown for a 100 (a) and 150\,$\mathrm{\mu}$m (b) thick sensor after an irradiation at 1$\times$10$^{16}$ $\mathrm{n}_{\mathrm{eq}}/\mathrm{cm}^2$.}
		\label{charge1e16}
	\end{figure}

	The collected charge is defined as the time integral of the induced current where, in this work, the integration time is chosen to be 10~ns. As depicted in Fig.~\ref{TCTsetup} the laser beam is moved in steps along the edge of the sensor. With this methodology, charge collection as a function of detector depth can be measured. In this section the results of 100\,$\mathrm{\mu}$m or 150\,$\mathrm{\mu}$m thick sensors are shown after an irradiation with neutrons in the TRIGA nuclear reactor of Jo\v{z}ef Stefan Institut \cite{reactor} up to a fluence of 1$\times$10$^{16}$ $\mathrm{n}_{\mathrm{eq}}/\mathrm{cm}^2$. After irradiation, the sensors are annealed for 80 min at 60$^{\circ}$C to complete the short term annealing of electrically active defects \cite{Lind}. In Fig.~\ref{charge1e16} the charge collection as a function of the sensor depth is summarised for voltages from 0 to 700\,V. Sensor front and back side are highlighted with the red dashed lines. The sensor front side is represented by the left line. The laser beam is directed to the middle of the pixel implant. The charge is listed in arbitrary units, meaning that the charges are comparable within one set of measurements with the same sample and experimental conditions but not across different samples. This is due to the fact that the absolute value of the current is dependent on the exact laser focus and power as well as the properties of the sensors edge surface. In Fig.~\ref{charge1e16}a the charge is collected through the entire detector thickness already at a bias voltage of 50\,V although the detector is irradiated to very high fluences of 1$\times$10$^{16}$ $\mathrm{n}_{\mathrm{eq}}/\mathrm{cm}^2$ and only a very thin depleted layer is expected at this bias voltage when assuming a constant effective doping concentration $N_{eff}$ \cite{Gregor2}. This is in agreement with the observation from Ref.~\cite{Gregor2} that at very high irradiation fluences the electric field is present in the entire detector volume already at low bias voltages. This is explained by space charge polarization and low free carrier concentration in highly irradiated material. The space charge polarization also results in the so called "double peak" effect \cite{doublepeak} visible in Fig.~\ref{charge1e16}b. The higher charge with respect to the central region is collected at the backside of the device at around 250\,V. 

\begin{figure}[h!]
	\centering     
	\subfigure[U=200\,V, d=100\,$\mathrm{\mu}$m]{\label{fig:a}\includegraphics[width=58mm]{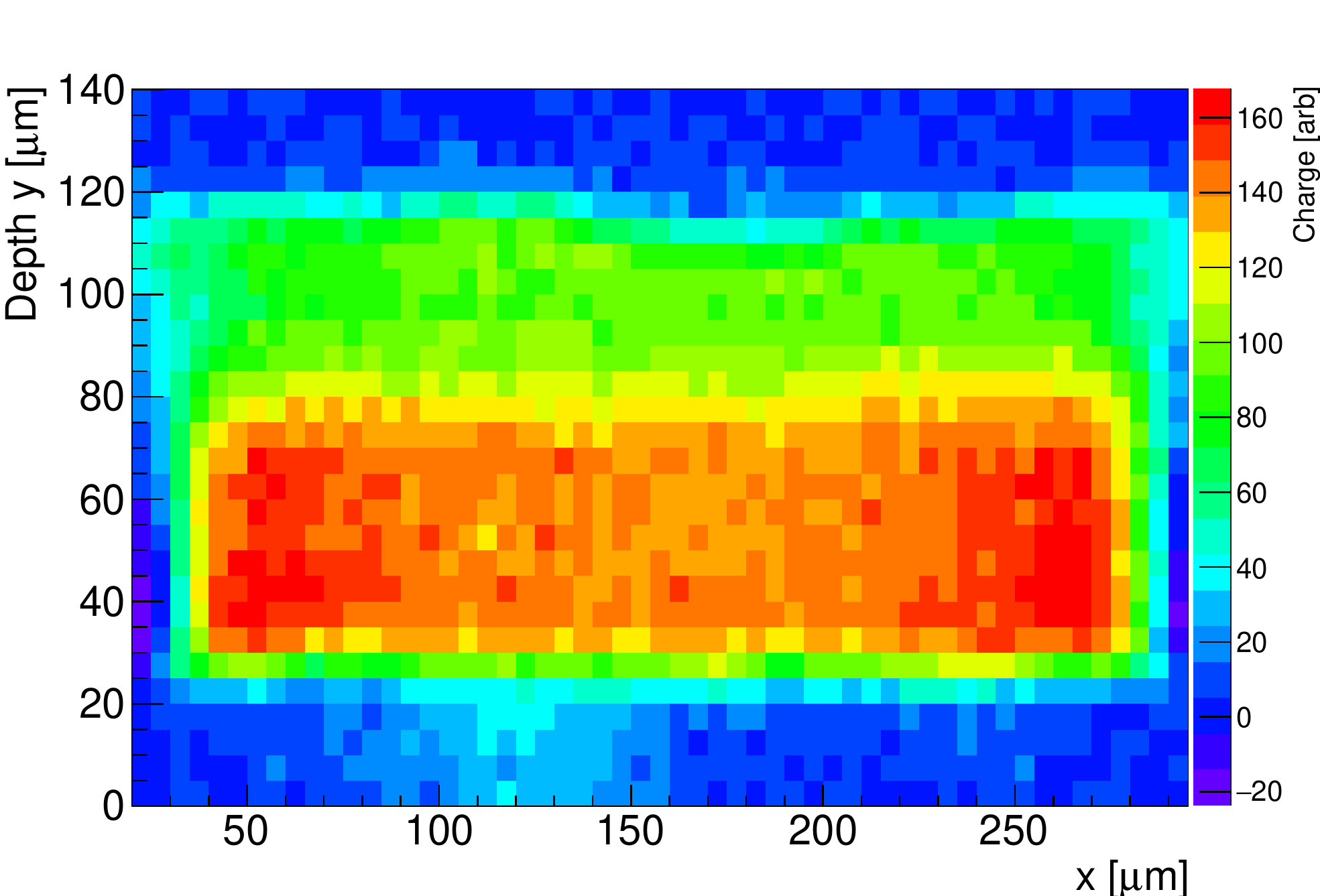}}
	\subfigure[U=600\,V, d=100\,$\mathrm{\mu}$m]{\label{fig:b}\includegraphics[width=58mm]{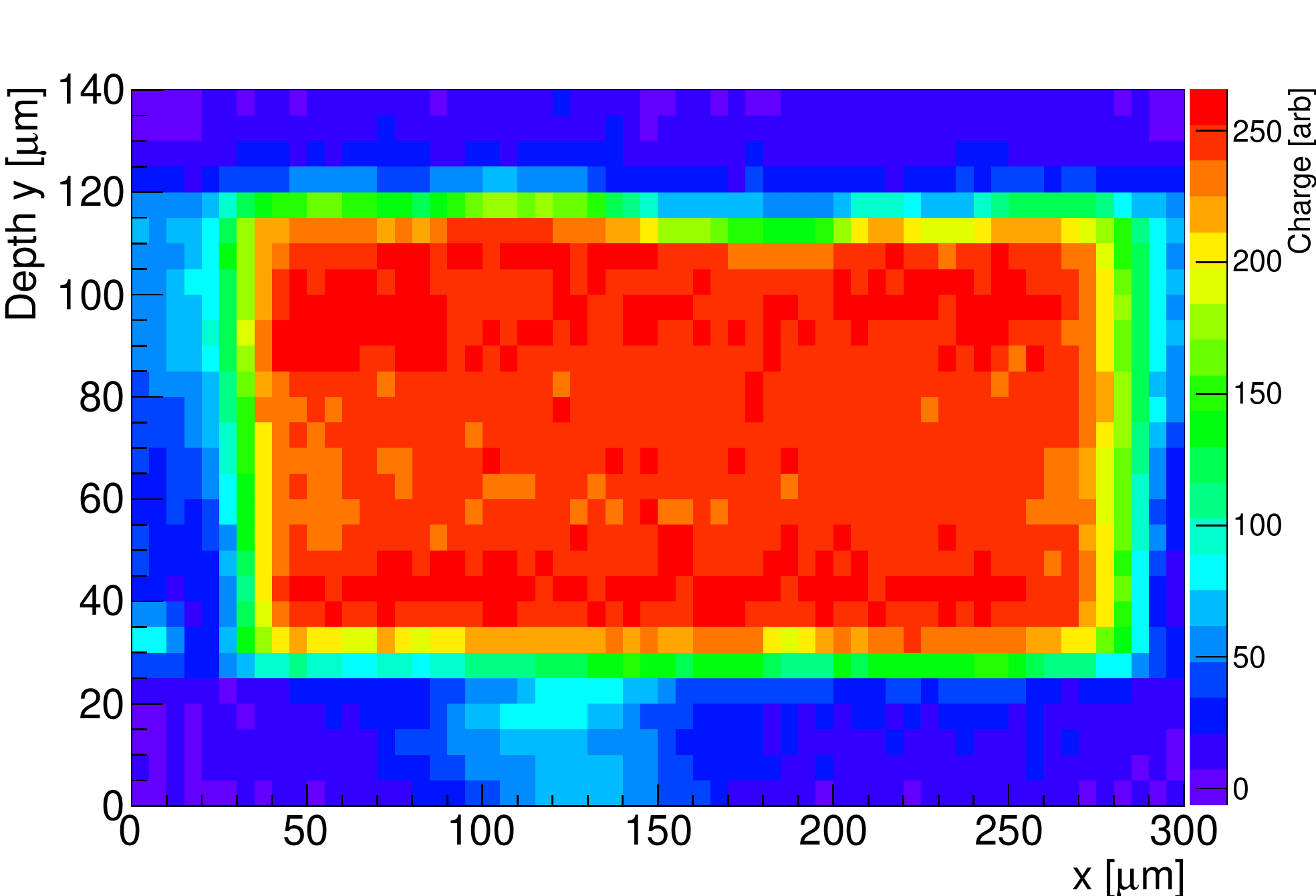}}
	\subfigure[U=200\,V, d=150\,$\mathrm{\mu}$m]{\label{fig:a}\includegraphics[width=58mm]{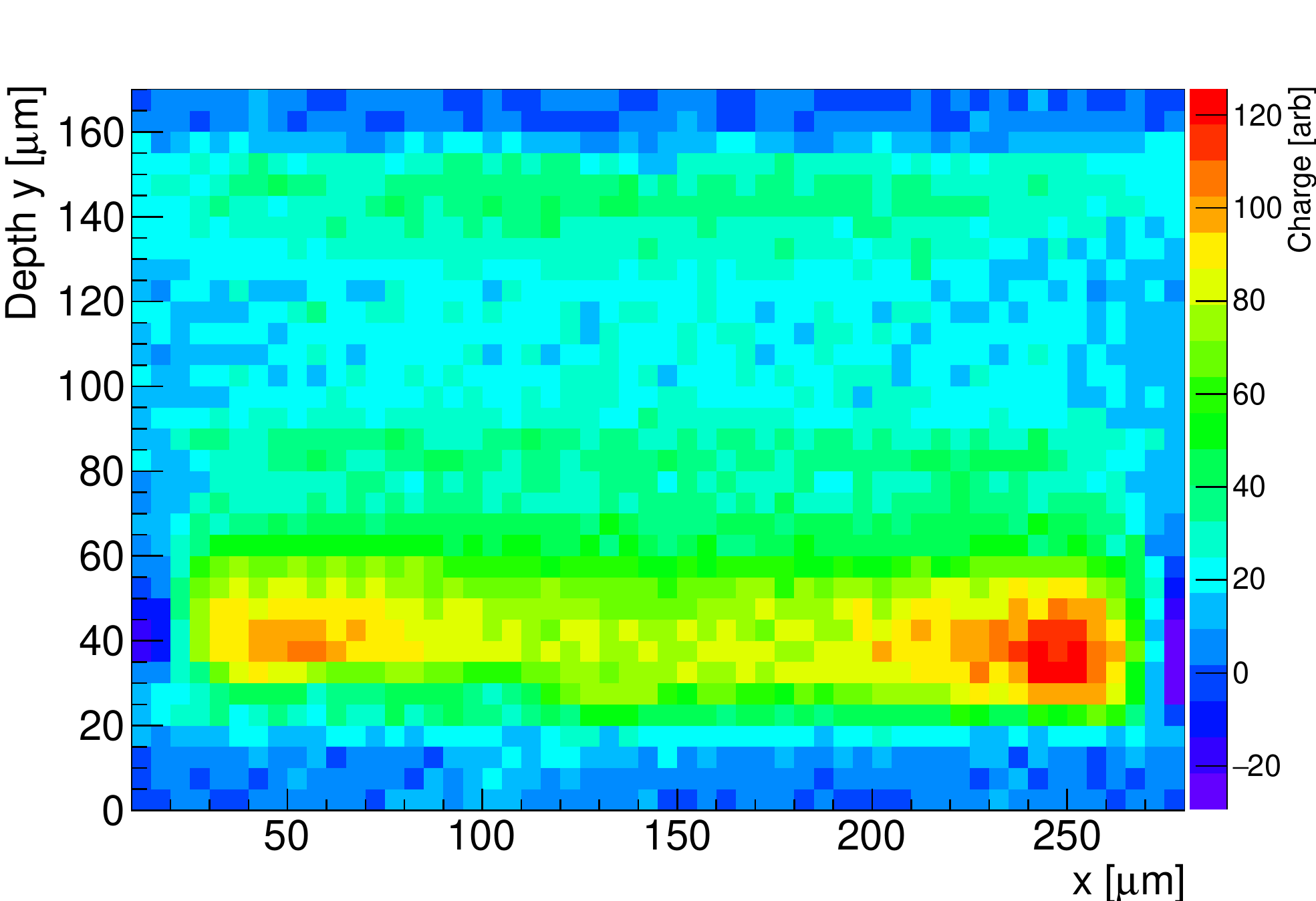}}
	\subfigure[U=600\,V, d=150\,$\mathrm{\mu}$m]{\label{fig:b}\includegraphics[width=58mm]{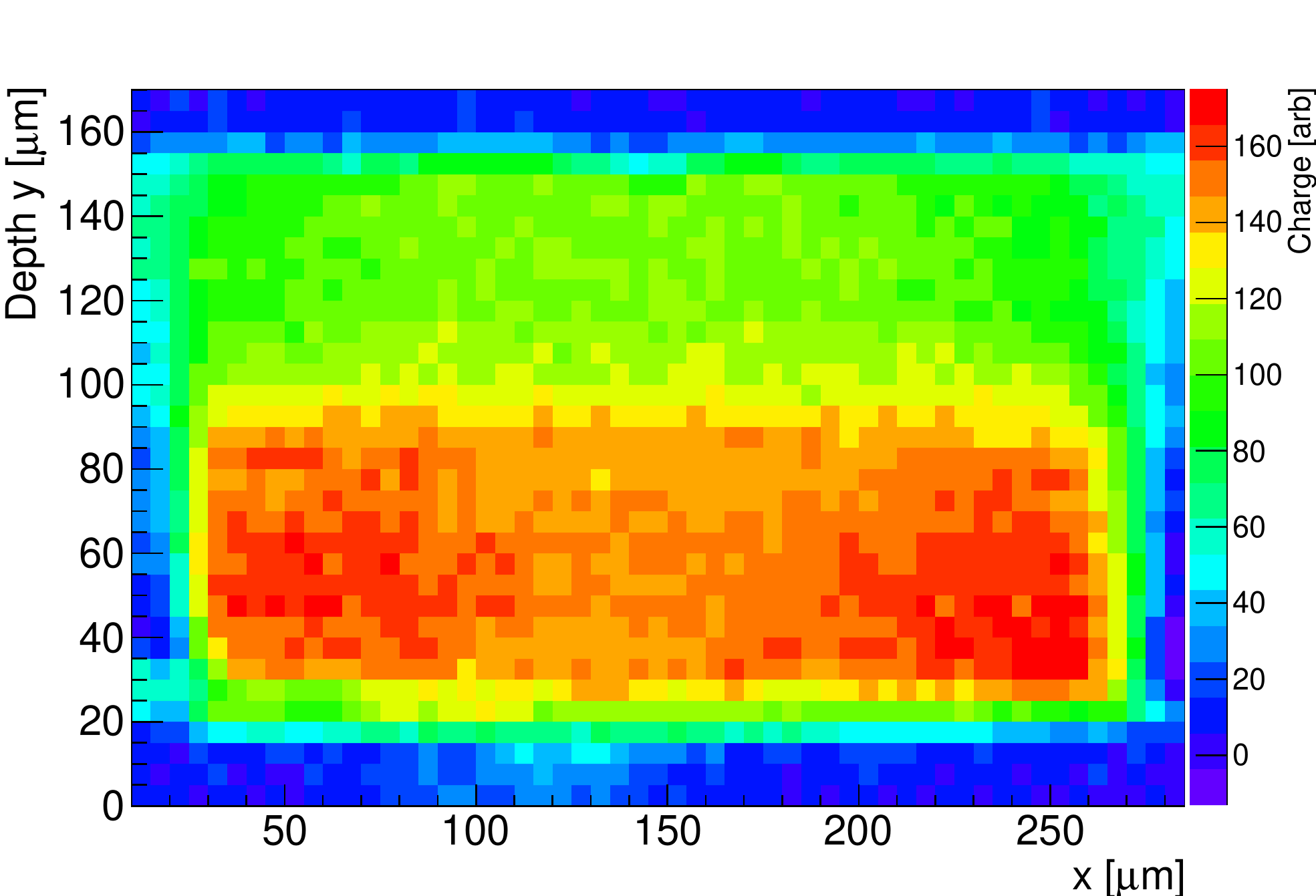}}
	\caption{Two-dimensional representation of a charge collection measurement with the Edge-TCT method for one single pixel cell in a 100\,$\mathrm{\mu}$m (a,b) or a 150\,$\mathrm{\mu}$m (c,d) thick sensor irradiated to 1$\times$10$^{16}$ $\mathrm{n}_{\mathrm{eq}}/\mathrm{cm}^2$, (a,c) at 200\,V and (b,d) at 600\,V. The collected charge is color-coded.}
	\label{2dcharge100}
\end{figure}

\noindent\hspace*{0mm}From 300\,V onwards the charge flattens throughout the sensor depth for the 100\,$\mathrm{\mu}$m device. For voltages above 450\,V no significant change of charge collection is observed, which points to a saturation of the drift velocity. Due to a lower electric field in thicker devices, this effect is not seen for the 150\,$\mathrm{\mu}$m thick sensor. It was not possible to apply higher voltages due to limited cooling capabilities. A two-dimensional representation of the charge collection measurement of the 100\,$\mathrm{\mu}$m and 150\,$\mathrm{\mu}$m thick sensors is depicted in Fig.~\ref{2dcharge100}. The laser beam direction is directed perpendicular to the X-Y plane. The sensor front side is at lower y-values with a pixel cell size of 250\,$\mathrm{\mu}$m in x. The depth of the sensor is along the y-axis with the detector surface being at about y=20\,$\mathrm{\mu}$m. Red areas represent high charges and the blue colour corresponds to low charges. The negative charge is a contribution from charges generated in the read-out pixel but collected from the adjacent one. The small charge measured at low y values (i.e when the laser beam is below the detector surface) is caused by light reflection of the wire bond on the read-out pixel. The high charge region at the sensor front side is increasing with voltage for both sensor thicknesses. Comparing both sensors reveals that in the 100\,$\mathrm{\mu}$m thin sensor the high charge region extends deeper both at 200 and 600\,V.

	\subsection{Electric field properties}
	According to Ramo's theorem \cite{Ramo}, the movement of charges inducing a current $I$ on the read-out pixel can be approximated in the case of a short laser pulse in the Edge-TCT configuration by:
	\begin{equation}
	I(y)= q \cdot N \cdot \vec v(y) / D
	\end{equation}
	where $q$ is the electric charge of the carrier, $N$ the number of carriers created at depth y, $\vec v(y)$ the carrier velocity and $D$ the detector thickness. The term $1/D$ represents the weighting field term for the Edge-TCT set-up with uniform charge generation underneath several adjacent pixels \cite{Gregor}. 
	
	\begin{figure}[h!]
		\centering     
		\subfigure[d=100\,$\mathrm{\mu}$m]{\label{}\includegraphics[width=65mm]{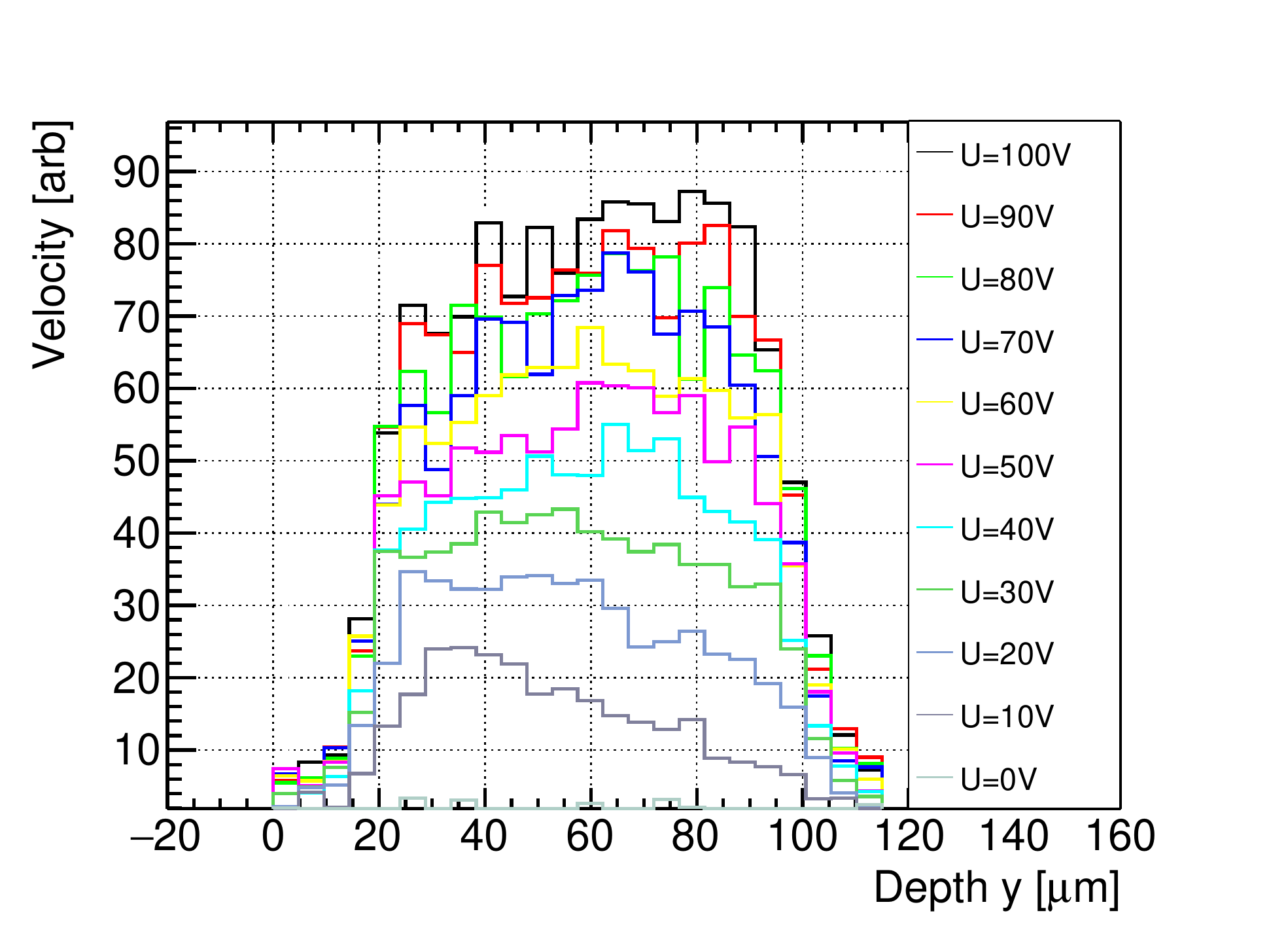}}
		\subfigure[d=150\,$\mathrm{\mu}$m]{\label{}\includegraphics[width=65mm]{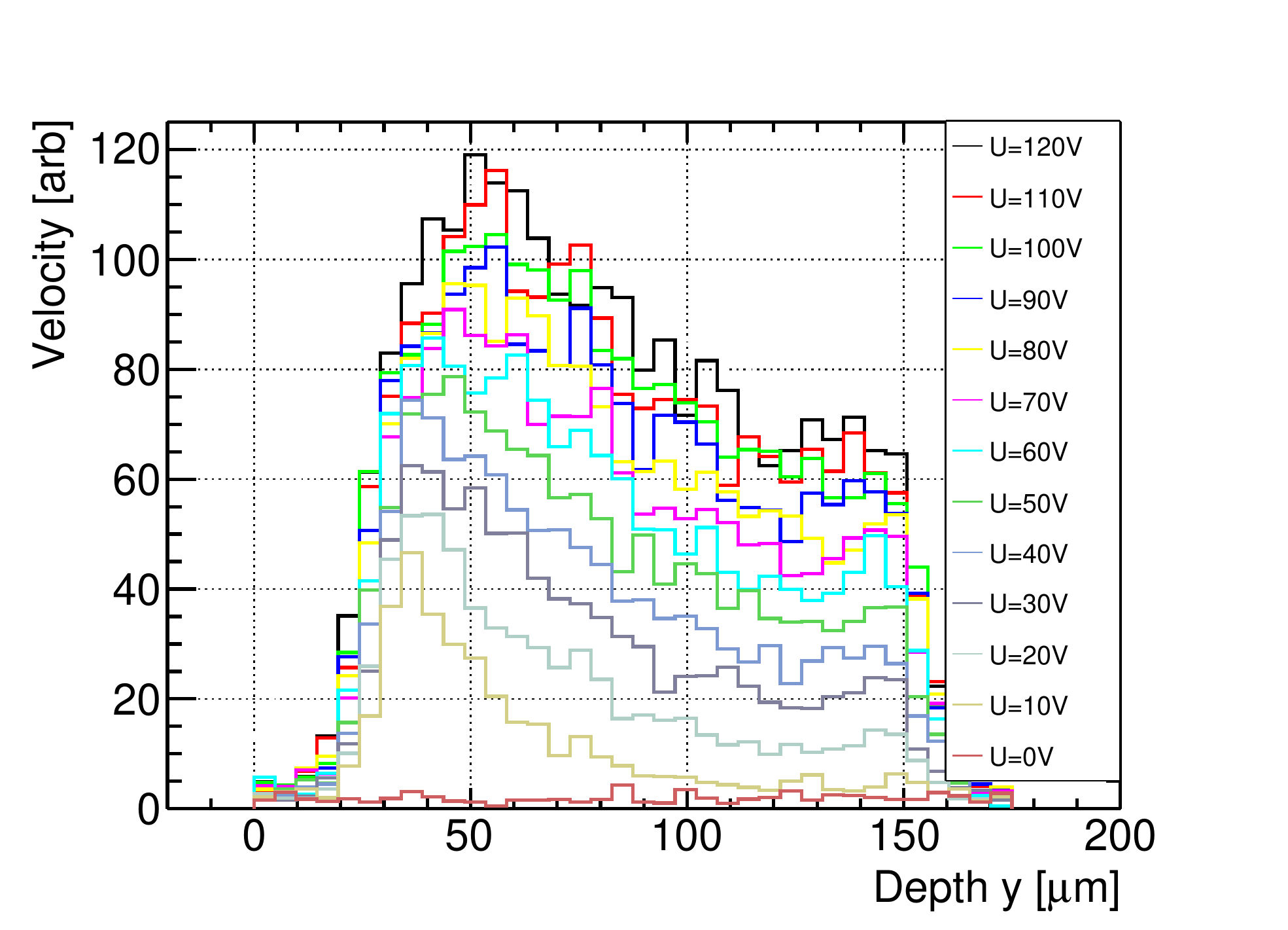}}
		\hfill
		\subfigure[d=100\,$\mathrm{\mu}$m, $\Phi$=5$\times$10$^{15}$ $\mathrm{n}_{\mathrm{eq}}/\mathrm{cm}^2$]{\label{}\includegraphics[width=65mm]{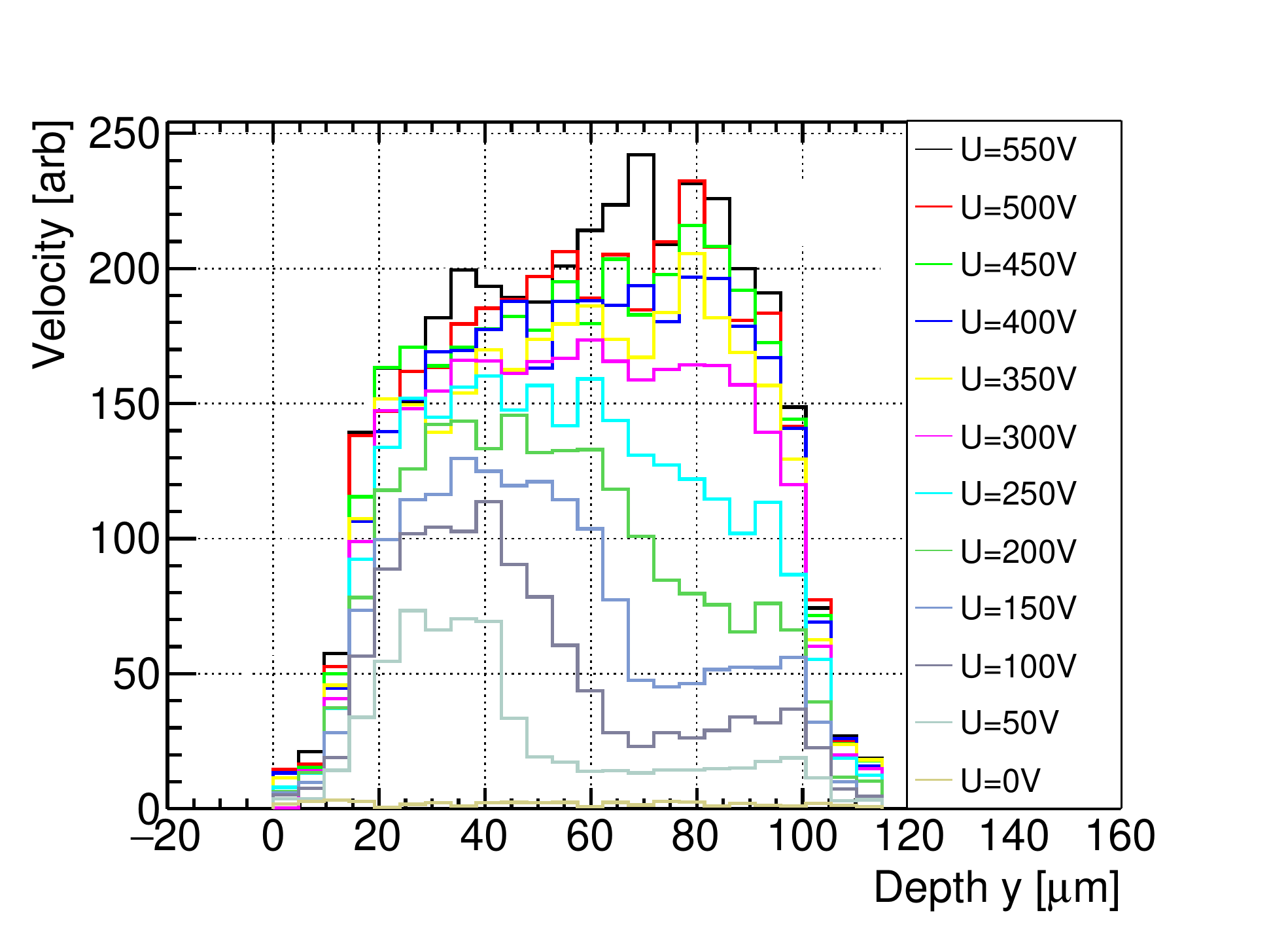}}
		\subfigure[d=100\,$\mathrm{\mu}$m, $\Phi$=1$\times$10$^{16}$ $\mathrm{n}_{\mathrm{eq}}/\mathrm{cm}^2$]{\label{}\includegraphics[width=65mm]{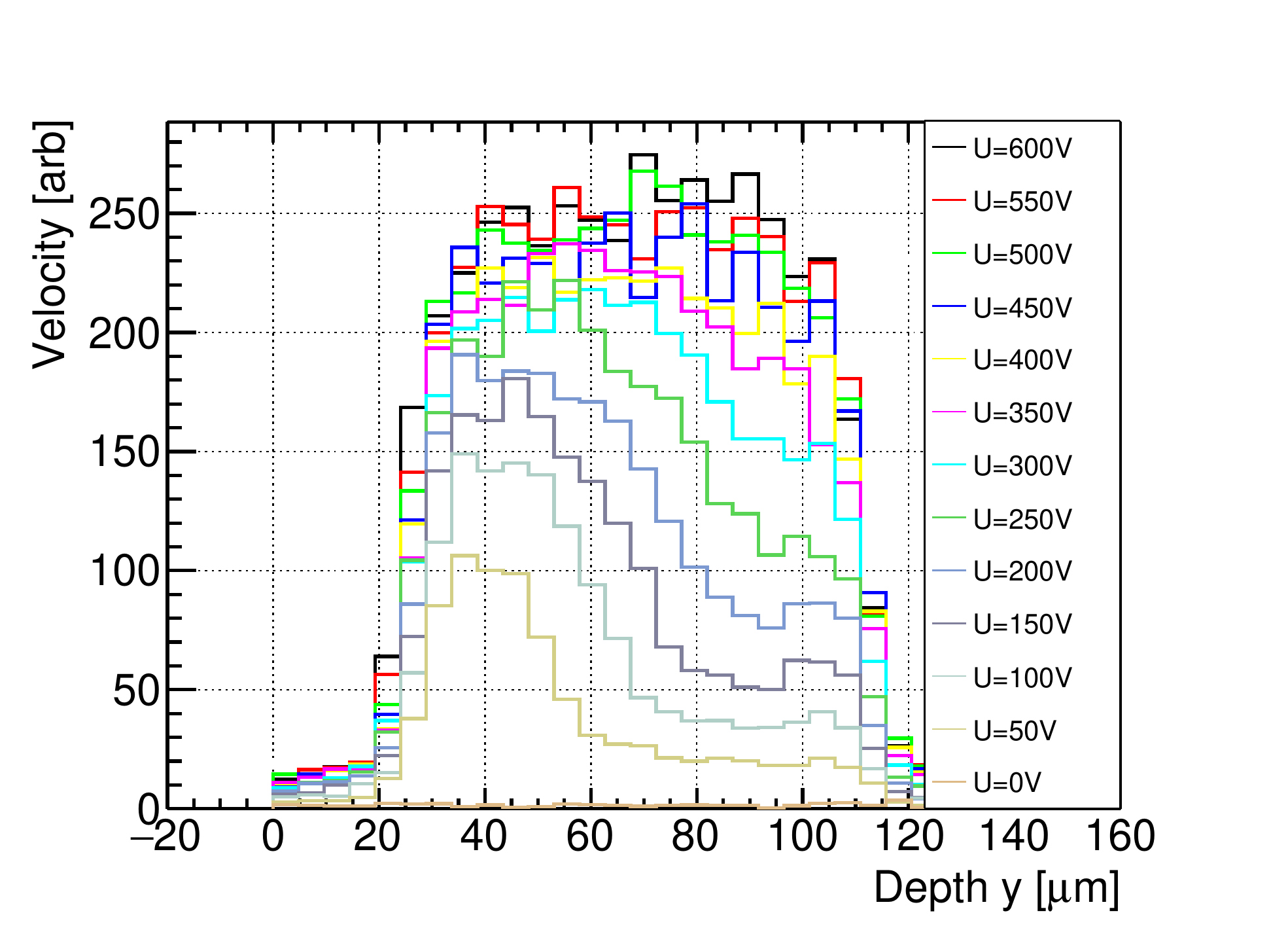}}
		\caption{Velocity profile as a function of sensor depth for not irradiated 100\,$\mathrm{\mu}$m (a) and 150\,$\mathrm{\mu}$m (b) thick sensors and for a 100\,$\mathrm{\mu}$m thick sensor after an irradiation to 5$\times$10$^{15}$ $\mathrm{n}_{\mathrm{eq}}/\mathrm{cm}^2$ (c) and 1$\times$10$^{16}$ $\mathrm{n}_{\mathrm{eq}}/\mathrm{cm}^2$ (d). The values of the velocity correspond to arbitrary units.}
		\label{velocity100and150}
	\end{figure}
	
\noindent\hspace*{0mm}If measured immediately after the pulse, before the carriers drift away from the location of the laser beam, $\vec v(y)$ is the velocity sum of the holes and electrons at depth $y$. Hence, it is possible to obtain the velocity profile from the increase of the signal in the very first ns. From the velocity profile the shape at the electric field can be inferred. In Fig.~\ref{velocity100and150} the velocity profiles of not irradiated 100\,$\mathrm{\mu}$m and 150\,$\mathrm{\mu}$m thick sensors are compared. A higher carrier velocity at the front side at lower voltages is observed for both thicknesses. At higher voltages the distribution is flatter for the 100\,$\mathrm{\mu}$m thick device, both before and after irradiation due to a more uniform electric field in thinner devices. The existence of the electric field at the back side of the sensor results in the devices to be fully depleted at 10\,V in case of the 100\,$\mathrm{\mu}$m detector and at 20\,V for the 150\,$\mathrm{\mu}$m thick detector. The velocity profiles after irradiation confirm the above hypothesis that an electric field in the entire sensor thickness already exists at low bias voltages. A double peak electric field distribution can be seen for both sensor thicknesses at low moderated bias voltage below 200\,V, but the field flattens at higher bias voltages. However, the electric field at the back is much weaker with respect to the electric field at the main junction of the pixels. The double peak effect after irradiation for thicker sensors has been reported in several publications \cite{doublepeak}.
	
	\subsection{Efficiency of 100 and 150\,$\mathrm{\mu}$m thick sensors after exposure to high irradiation}
\begin{figure}[htpb!]
		\centering     
		\includegraphics[width=60mm]{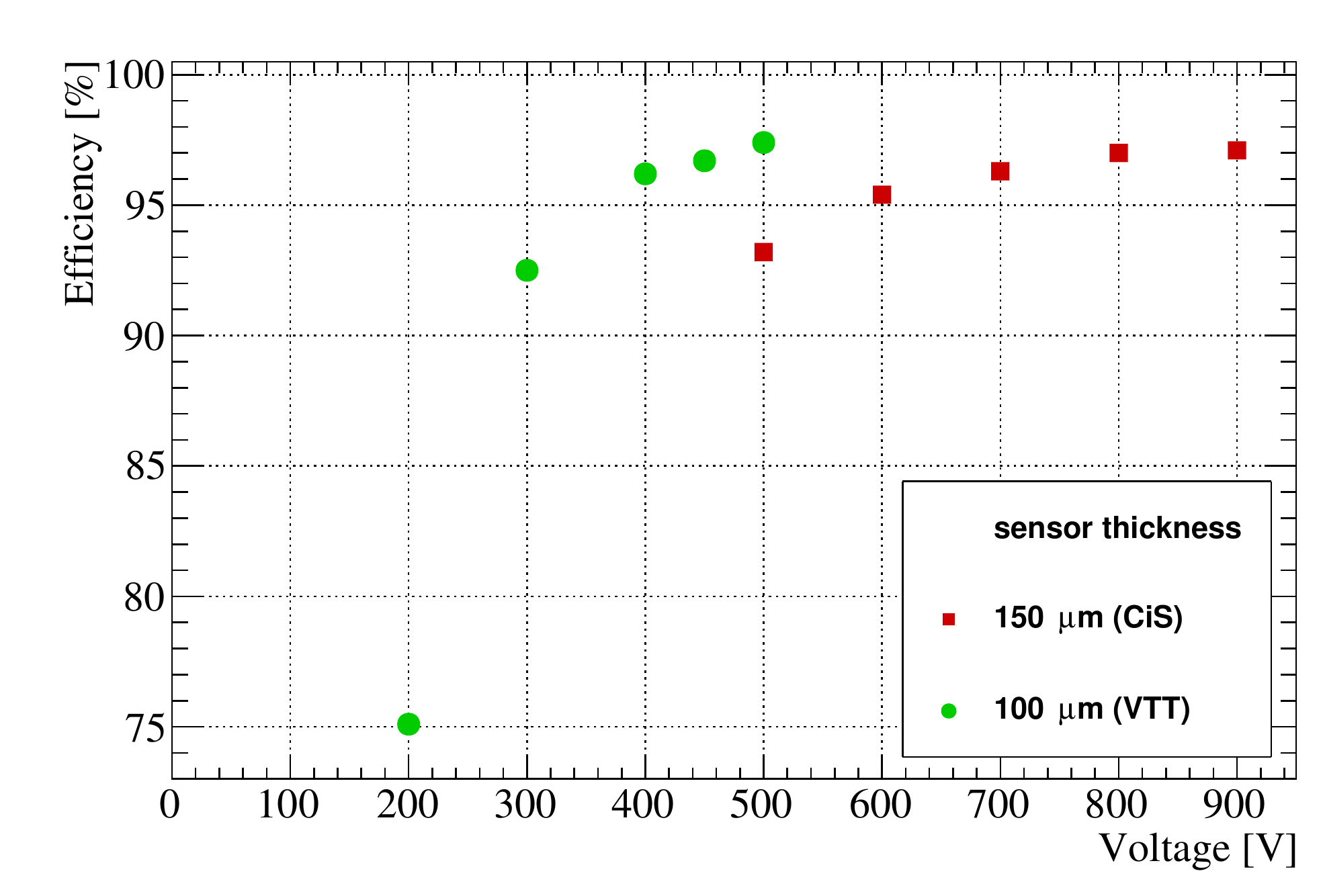}
		\caption{Comparison of hit efficiencies of FE-I4 modules with 100\,$\mathrm{\mu}$m or 150\,$\mathrm{\mu}$m thick sensors irradiated to 1$\times$10$^{16}$ $\mathrm{n}_{\mathrm{eq}}/\mathrm{cm}^2$ at Jo\v{z}ef-Stefan Institute in Ljubljana and the PS at CERN.}
		\label{eff1e16}
	\end{figure}

	An FE-I4 module employing the 150\,$\mathrm{\mu}$m thick sensor irradiated to a fluence of 1$\times$10$^{16}$ $\mathrm{n}_{\mathrm{eq}}/\mathrm{cm}^2$ is measured with particle beams at CERN. The results are compared to those of an FE-I4 module with a 100\,$\mathrm{\mu}$m thick sensor previously reported in \cite{nani2}. In Fig.~\ref{eff1e16} for 500\,V the hit efficiency of the 100\,$\mathrm{\mu}$m device of 97.4\% is higher than the efficiency of the 150\,$\mathrm{\mu}$m sensor. The efficiency of the module with a 150\,$\mathrm{\mu}$m thick sensor saturates at a similiar value of 97.0\% with the significant difference being the higher bias voltage of around 800\,V needed to achieve the saturation. These results support the charge collection results obtained from Edge-TCT measurements and demonstrate the feasibility of instrumenting the innermost pixel layers at HL-LHC with 100\,$\mathrm{\mu}$m thin planar sensors.
	
	\section{Performance of irradiated sensors with small pixel cells}
\begin{figure}[h!]
	\centering     
	\includegraphics[width=55mm]{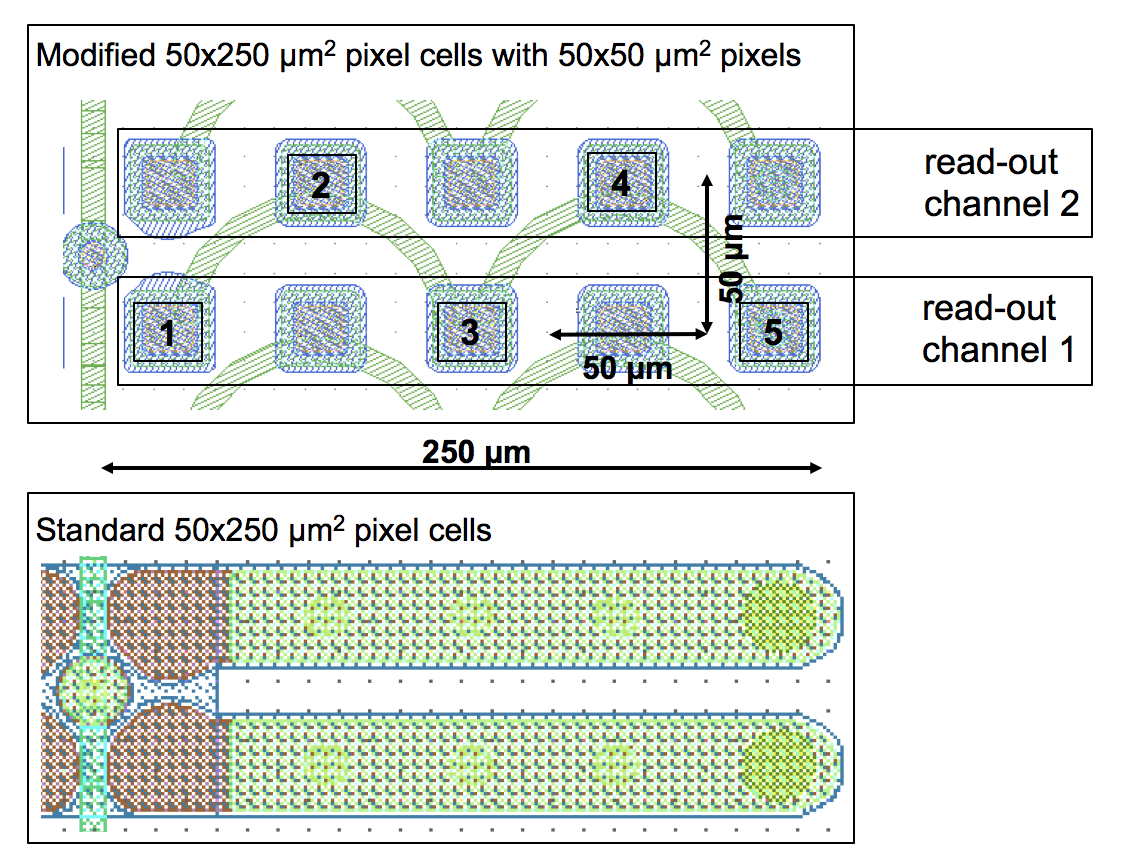}
	\caption{A modified 50x250\,$\mathrm{\mu}$m$^{2}$ pixel cell composed of five 30x30\,$\mathrm{\mu}$m$^{2}$ pixel implants with a 50\,$\mathrm{\mu}$m pixel pitch in both directions is shown and compared to a standard pixel cell. The implants are connected by metal lines allowing for readout by the FE-I4 chip. In both structures the common punch-through is implemented.}
	\label{50x50design}
\end{figure}

\begin{figure}[h!]
	\centering     
	\subfigure[]{\label{}\includegraphics[height=4.5cm]{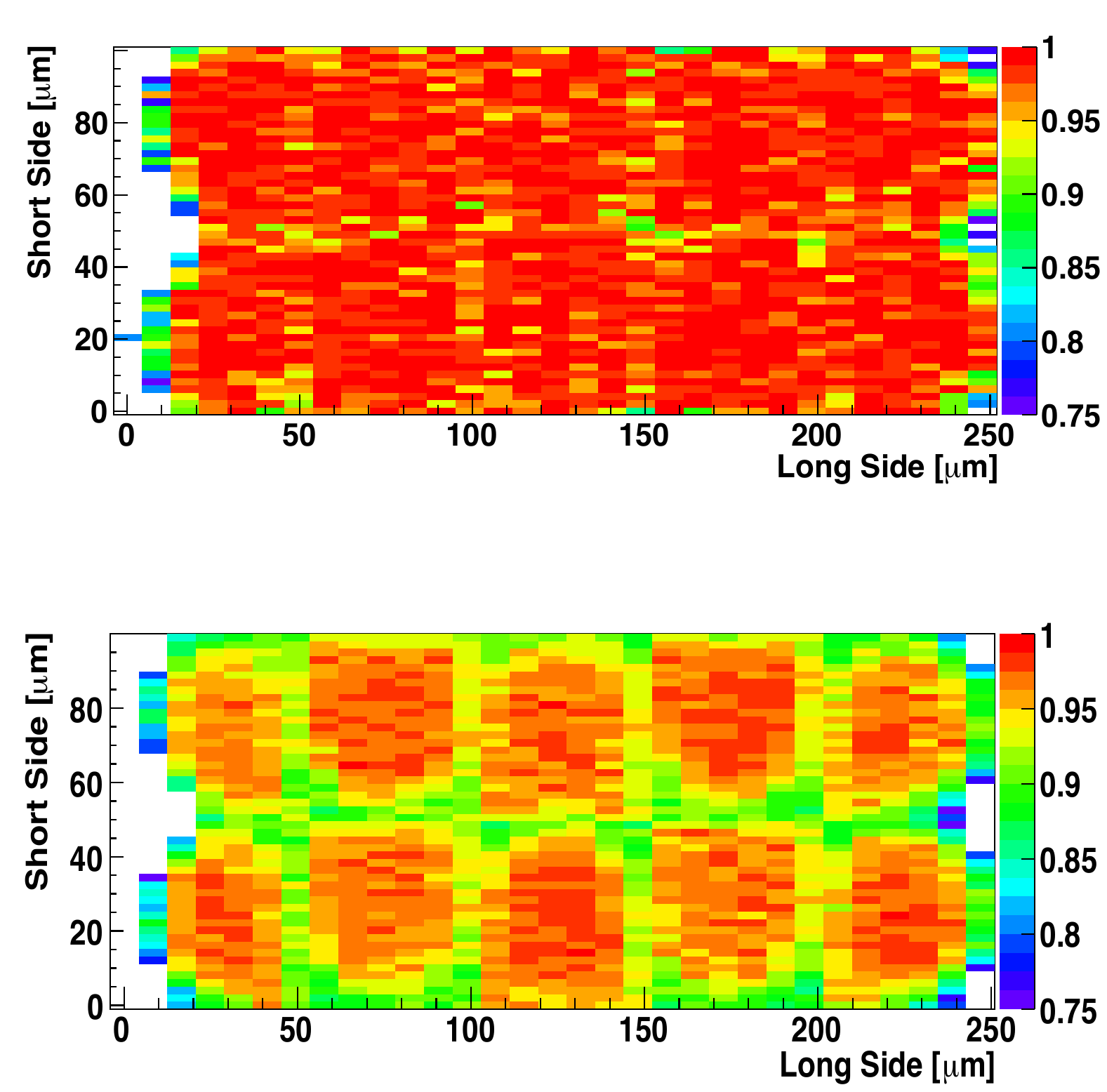}}
	\subfigure[]{\label{}\includegraphics[height=4.5cm]{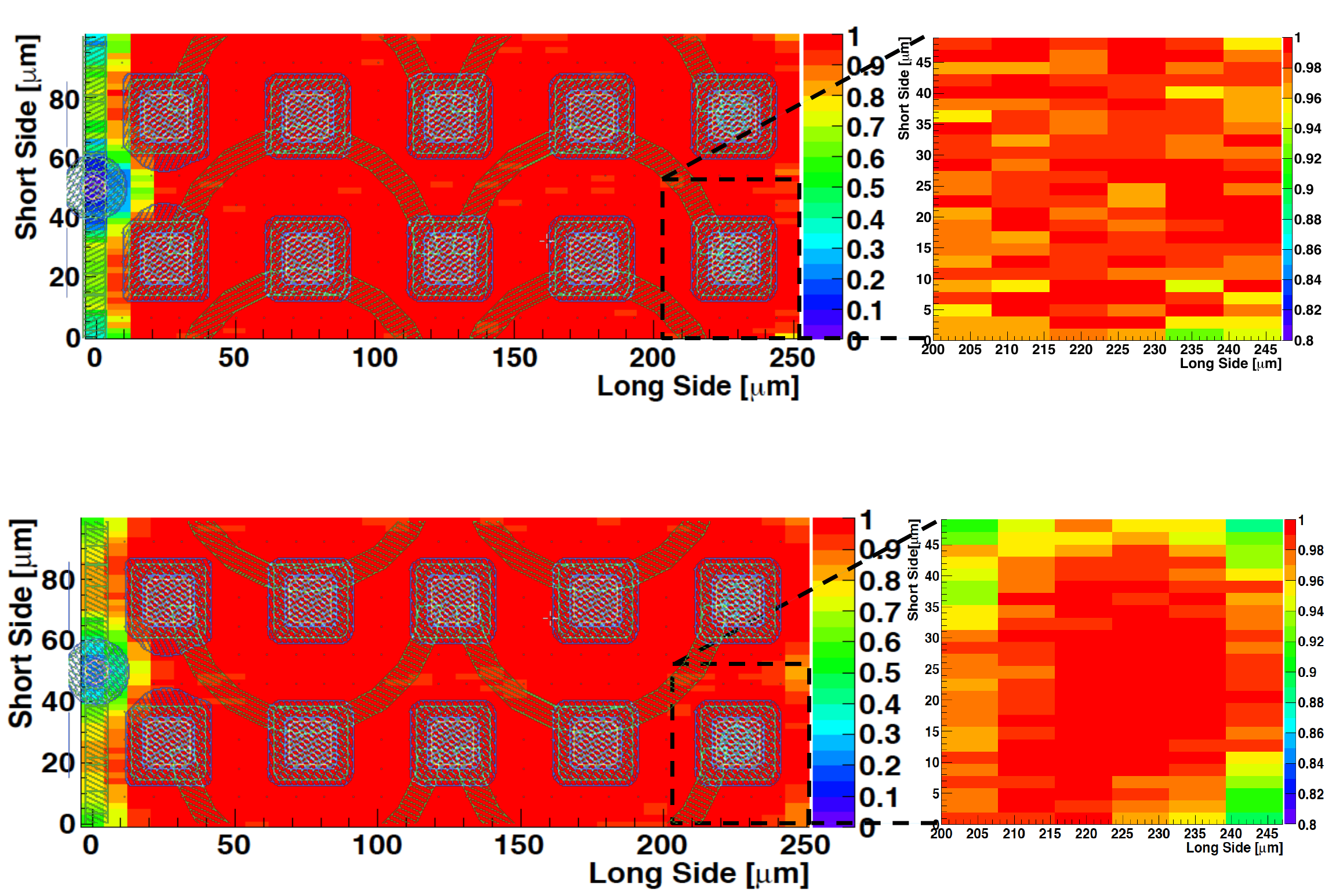}}
	\caption{In-pixel efficiency map for a 150\,$\mathrm{\mu}$m thin sensor (top) at 300\,V (a) and 500\,V (b) irradiated to 3$\times$10$^{15}$ $\mathrm{n}_{\mathrm{eq}}/\mathrm{cm}^2$ and a 100\,$\mathrm{\mu}$m thin sensor (bottom) at 500\,V (a) and 600\,V (b) irradiated to 5$\times$10$^{15}$ $\mathrm{n}_{\mathrm{eq}}/\mathrm{cm}^2$. In (a) the scale is zoomed to enhance the efficiency loss in between the pixel cells. In (b) the design is overlayed and the efficiency of the edge implant is calculated for the closest approximation of the geometry of a RD53A 50x50\,$\mathrm{\mu}$m$^{2}$ pixel cell without punch-through structure.}
	\label{50x50eff}
\end{figure}

For smaller pixels cells of 50x50\,$\mathrm{\mu}$m$^{2}$ a more prominent effect of charge sharing is expected compared to larger pixel cells. A modified 50x250\,$\mathrm{\mu}$m$^{2}$ pixel cell composed of five 30x30~$\mathrm{\mu}$m$^{2}$ pixel implants with a 50\,$\mathrm{\mu}$m pixel pitch in both directions was designed to investigate the charge sharing in adjacent 50x50\,$\mathrm{\mu}$m$^{2}$ pixels. The implants are connected by metal lines allowing for the readout by the FE-I4 chip. Neigbouring pixel implants are read-out by different channels. A sketch of the design of the modified pixel cell, and a comparison to the standard pixel cell, are displayed in Fig. \ref{50x50design}. Before irradiation, no significant charge sharing among neighbouring pixels is observed although the charge would be splitted between two adjacent pixel readout channels \cite{nani2}. This is due to the fact that the charge in the pixel cell is still above threshold, thus not affecting the efficiency. After irradiation the effect of charge sharing for smaller pixel cells is evident at lower bias voltages. The in-pixel efficiency maps after an irradiation to 3 and 5$\times$10$^{15}$ $\mathrm{n}_{\mathrm{eq}}/\mathrm{cm}^2$ are shown in Fig.~\ref{50x50eff} for the sensors biased at 300\,V and 500\,V (a) and at 500\,V and 600\,V (b). The highlighted pixel cell is chosen to be the closest approximation to the geometry of RD53A compatible pixel cells without punch-through structure. It is surrounded by eight pixel cells, where seven of them are readout by different readout channels, while only one is read out by the same channel. The efficiency of the edge 50x50~$\mathrm{\mu}$m$^{2}$ pixel cell is calculated to be 98.2\% at 500\,V for a 150\,$\mathrm{\mu}$m sensor irradiated to 3$\times$10$^{15}$ $\mathrm{n}_{\mathrm{eq}}/\mathrm{cm}^2$, while an efficiency of 97.5\% is achieved at 600\,V for a 100\,$\mathrm{\mu}$m sensor irradiated to 5$\times$10$^{15}$ $\mathrm{n}_{\mathrm{eq}}/\mathrm{cm}^2$. With this modified geometry of the FE-I4 compatible sensor design it is therefore possible to make first predictions on the performance of 50x50\,$\mathrm{\mu}$m$^{2}$ pixel cells after irradiation.

	\section{Conclusions}
	At CiS 100 and 150~$\mathrm{\mu}$m thick sensors with standard and modified 50x250~$\mathrm{\mu}$m$^{2}$ pixel cells were produced. The thin sensors are characterized by Edge-TCT and beam test measurements. Charge collection and velocity are extracted from Edge-TCT measurements and compared to efficiency results obtained during beam test campaigns. For a 100\,$\mathrm{\mu}$m thick sensor the charge collection profiles at all fluences flatten above 350-400\,V. The 150\,$\mathrm{\mu}$m thick irradiated sensor does not reach this condition given the restriction in the bias voltage caused by limitations of the available cooling system. In the chosen range of bias voltages the electric field is flatter for the 100\,$\mathrm{\mu}$m thick sensors with respect to the 150\,$\mathrm{\mu}$m thick sensors. An indication of a double peak in the electric field can only be observed at low bias voltages. Results obtained with particle beams support the Edge-TCT results. A module with a  100\,$\mathrm{\mu}$m thick sensor reaches an efficiency of 97.4\% at a moderate bias voltage of 500\,V. These results show that this technology can be applied in the innermost layers of the HL-LHC pixel systems. Furthermore, modified 50x250\,$\mathrm{\mu}$m$^{2}$ pixel cells with 30x30\,$\mathrm{\mu}$m$^{2}$ pixel implants were produced to investigate the performance of the future small pixel cells for the RD53A chip. A 50x50\,$\mathrm{\mu}$m$^{2}$ pixel cell in a 150\,$\mathrm{\mu}$m thick sensor after an irradiation to 3$\times$10$^{15}$ $\mathrm{n}_{\mathrm{eq}}/\mathrm{cm}^2$ is shown to achieve an efficiency of 98.2\%, and in the case of a 100\,$\mathrm{\mu}$m thick sensor irradiated to 5$\times$10$^{15}$ $\mathrm{n}_{\mathrm{eq}}/\mathrm{cm}^2$ an efficiency of 97.5\% is observed.
	
	\acknowledgments
	
	This work has been partially performed in the framework of the CERN RD50 Collaboration. The authors thank F. Ravotti for the irradiation at CERN-PS, I. Mandi\'{c} for the one at JSI, L. Gonella for the one at UoB. Supported by the H2020 project AIDA-2020, GA no. 654168. 
	

\end{document}